\documentclass[]{aa}
\usepackage{amsmath}
\usepackage[varg]{txfonts}
\usepackage{natbib}
\bibpunct{(}{)}{;}{a}{}{,} 
\usepackage{graphicx}
\usepackage{placeins}

\authorrunning{S. Th\"olken et al.}

\title{XMM-Newton X-ray and HST weak gravitational lensing study of the extremely X-ray luminous
galaxy cluster \RXJ (\mbox{$z=0.902$})}

\author{Sophia Th\"olken\inst{\ref{inst1}}\
\and Tim Schrabback\inst{\ref{inst1}}\
\and Thomas H. Reiprich\inst{\ref{inst1}}\
\and Lorenzo Lovisari\inst{\ref{inst2}}\
\and Steven W. Allen\inst{\ref{inst3},\ref{inst4},\ref{inst5}}\
\and Henk Hoekstra\inst{\ref{inst6}}\
\and Douglas Applegate \inst{\ref{inst7}}\
\and Axel Buddendiek\inst{\ref{inst1}}\
\and Amalia Hicks\inst{\ref{inst8}}
}

\authorrunning{S. Th\"olken et al.}
\titlerunning{XMM-Newton X-ray and HST weak gravitational lensing study of \RXJ (\mbox{$z=0.902$})}

\institute{Argelander-Institut f\"ur Astronomie, Universit\"at Bonn, Auf dem H\"ugel 71, 53121 Bonn, Germany\label{inst1}\\
\email{thoelken@astro.uni-bonn.de}
\and
Harvard-Smithsonian Center for Astrophysics, 60 Garden Street, Cambridge, MA 02138, USA\label{inst2}
\and
Kavli Institute for Particle Astrophysics and Cosmology, Stanford University, 452 Lomita Mall, Stanford, CA 94305-4085, USA\label{inst3}
\and
Department of Physics, Stanford University, 452 Lomita Mall, Stanford, CA 94305-4085, USA\label{inst4}
\and
SLAC National Accelerator Laboratory, 2575 Sand Hill Road, Menlo Park, CA 94025, USA\label{inst5}
\and
Leiden Observatory, Leiden University, PO Box 9513, 2300 RA Leiden, the Netherlands\label{inst6}
\and
Kavli Institute for Cosmological Physics, University of Chicago, 5640 S Ellis Ave, Chicago, IL 60637\label{inst7}
\and 
Cadmus, Energy Services Division, ​16 N. Carroll Street, Suite 900, Madison, WI 53703\label{inst8}
}
\def\RXJ{Cl\thinspace$J$120958.9+495352 }
\def\RXJdot{Cl\thinspace$J$120958.9+495352}
\newcommand{\clustera}{Cl\thinspace$J$120958.9+495352\,}
\newcommand{\clusteranospace}{Cl\thinspace$J$120958.9+495352}
\date{Received date /
Accepted date }

\abstract {Observations of relaxed, massive and distant clusters can provide important tests of standard cosmological models, for example by using the gas mass fraction. To perform this test, the dynamical state of the cluster and its gas properties have to be investigated. X-ray analyses provide one of the best opportunities to access this information and to determine important properties such as temperature profiles, gas mass, and the total X-ray hydrostatic mass. For the last of these, weak gravitational lensing analyses are complementary independent probes that are essential in order to test whether X-ray masses could be biased.} 
{We study the very luminous, high redshift ($z=0.902$) galaxy cluster \RXJ using XMM-Newton data. We measure global cluster properties and study the temperature profile and the cooling time to investigate the dynamical status with respect to the presence of a cool core. We use Hubble Space Telescope (HST) weak lensing data to estimate its total mass and determine the gas mass fraction.} 
{We perform a spectral analysis using an XMM-Newton observation of 15\,ks cleaned exposure time. As the treatment of the background is crucial, we use two different approaches to account for the background emission to verify our results. We account for point spread function effects and deproject our results to estimate the gas mass fraction of the cluster. 
We measure weak lensing galaxy shapes from mosaic HST imaging and select background galaxies photometrically in combination with imaging data from the William Herschel Telescope.} 
{The X-ray luminosity of \RXJ in the $0.1-2.4$\,keV band estimated from our XMM-Newton data is $L_X = (13.4_{-1.0}^{+1.2})\times10^{44}$\,erg/s and thus
it is one of the most X-ray luminous clusters known at similarly high redshift. We find clear indications for the presence of a cool core from the temperature profile and the central cooling time, which is very rare at such high redshifts. Based on the weak lensing analysis, we estimate a cluster mass of \mbox{$M_\mathrm{500}/10^{14}M_\odot=4.4^{+2.2}_{-2.0}(\mathrm{stat.})\pm0.6(\mathrm{sys.})$} and a gas mass fraction of $f_{\rm gas,2500} = 0.11_{-0.03}^{+0.06}$ in good agreement with previous findings for high redshift and local clusters.} {}


\def\Vhrulefill{\leavevmode\leaders\hrule height 0.7ex depth \dimexpr0.4pt-0.7ex\hfill\kern0pt}

\begin{document}
\defcitealias{schrabback16}{S16}
\maketitle

\section{Introduction}
In the paradigm of hierarchical structure formation very massive and distant clusters should be extremely rare. These clusters provide the opportunity for many interesting astrophysical and cosmological studies. The gas mass fraction ($f_{\rm gas}$) of dynamically relaxed clusters is an important probe of cosmological models (\citealp{2008MNRAS.383..879A}, \citealp{2014MNRAS.440.2077M}) as the matter content of these objects should approximately match the matter content of the universe (e.g., \citealp{1993Natur.366..429W}, \citealp{2011ARA&A..49..409A}, and references therein). In particular high-redshift clusters are of interest where the leverage on the cosmology is largest.

The cooling time for these clusters is very short and the presence of a cool core is believed to be strongly related to the dynamical status of the cluster (e.g., \citealp{2010A&A...513A..37H}). \citet{2017arXiv170205094M} studied the evolution of the ICM and cool core clusters over the past 10\,Gyr. Their results imply that from redshift $z=0$ to $z=1.2$ cool cores basically do not evolve in size, density, and mass.
Additionally, the level of agreement of the properties of these rare clusters with existing scaling relations (e.g., \citealp{2011A&A...535A...4R}, \citealp{2009A&A...498..361P}) has great significance for cosmology as the properties can provide tests of these scaling laws and assess whether they are in line with standard cosmological predictions. 

So far, only a few of these rare, relaxed, massive, and high redshift objects have been found; two examples are Cl$J$0046.3+8530 (\citealp{2004MNRAS.354....1M}) and Cl$J$1226.9+3332 (\citealp{2004MNRAS.351.1193M}). 
In the Massive Cluster Survey (MACS) (\citealp{2007ApJ...661L..33E}, \citealp{2010MNRAS.407...83E}), many interesting objects have been identified, for example extreme cooling in cluster cores such as {MACS$J$1931.8-2634 (\citealp{2011MNRAS.411.1641E})}, and a number of dynamically relaxed clusters that can be used for cosmological tests. However, almost all of those relaxed clusters are at smaller redshifts than the object studied here. Two of the most distant clusters at $z>1$, Cl$J$1415.1+3612 ($z=1.028$) and 3C~186 ($z=1.067$), were studied in detail by \citet{2014BaltA..23...93B} and \citet{2010ApJ...722..102S} using deep Chandra observations. The observations revealed a cool core for both objects with a short cooling time for Cl$J$1415.1+3612 within the core region of $<0.2$\,Gyr and a gas-mass fraction consistent with local clusters for 3C 186. With respect to the luminosity, another extreme example is the El Gordo galaxy cluster at $z=0.87$ with $L_X = (2.19 \pm 0.11) \times 10^{45}\,h^{-2}_{70}$\,erg/s (\citealp{2012ApJ...748....7M}) which is one of the most massive and luminous clusters found so far.

\begin{figure}[b]
\resizebox{\hsize}{!}{\includegraphics{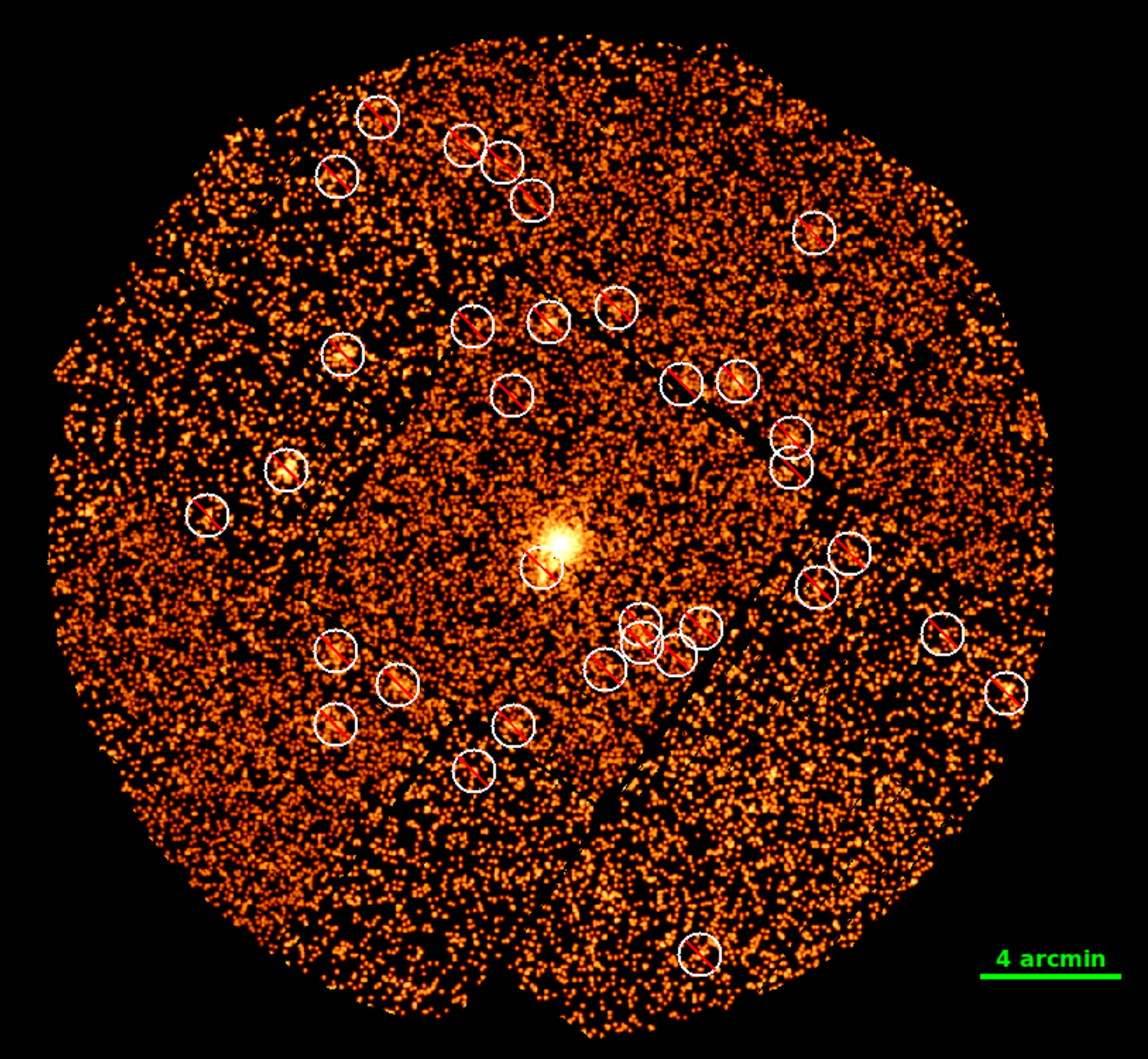}}
\caption{Combined, cleaned, exposure corrected, and smoothed MOS image of \RXJdot. White circles show the excluded point sources.}
\label{fig:RXJ}
\end{figure}

For cosmological tests, the total cluster mass is an important quantity for which weak gravitational lensing provides an independent probe in addition to the X-ray hydrostatic mass. The gravitational potential imprints coherent distortions
onto the observed shapes of background galaxies \citep[e.g.,][]{bartelmann01,schneider06}. 
Measurements of these weak lensing distortions directly
constrain the projected mass distributions and cluster masses
\citep[][]{hoekstra13}. 
These measurements are sensitive to the total
matter distribution, including  both dark matter and baryons. Especially at high redshifts, the Hubble Space Telescope (HST) is an essential tool for the analysis of such objects as ground-based telescopes are not able to resolve the shapes of the very distant background galaxies.

Recently, \citet{2015MNRAS.450.4248B} performed a combined search of distant massive clusters using ROSAT All Sky Survey and Sloan Digital Sky Survey (SDSS) data covering an area of 10,000\,deg$^2$. They found 83 high-grade candidates for X-ray luminous clusters between $0.6<z<1$ and obtained \textit{William Herschel Telescope} (WHT) or Large Binocular Telescope (LBT) imaging to confirm the candidates. One of the clusters they found is special in many respects: \RXJ is the most X-ray luminous cluster in their sample. Also, it has the second highest spectroscopically confirmed redshift in their sample, and their richness and Sunyaev-Zel'dovich (SZ) measurements independently indicate a high cluster mass. 
According to the Planck catalog of SZ sources (\citealp{2015A&A...581A..14P}) \RXJ is on par with the five most {X-ray} luminous clusters found at $z{\sim}0.9$. It is thus a valuable candidate for a distant cooling-core cluster and provides a great opportunity to study one of these rare systems in detail.

In this work we perform a spectroscopic XMM-Newton and HST weak lensing study of this extraordinary object found by \citet{2015MNRAS.450.4248B}. We investigate the temperature profile with respect to the presence of a cool core and determine the cooling time within $<100$\,kpc. In Sec. \ref{sec:obs} we describe the properties of \RXJdot, the data reduction procedure, and the analysis strategy for HST and XMM-Newton and for the XMM-Newton background. Sec. \ref{sec:results} gives the results, which are then discussed in Sec. \ref{sec:discussion}.

Throughout the analysis we use a flat $\Lambda$CDM cosmology with $H_0 = 70$\,km/s/Mpc, $\Omega_m=0.3$ and $\Omega_\Lambda = 0.7$. All uncertainties are given at the 68\% confidence level and overdensities refer to the critical density. All magnitudes are in the AB system.

\section{Observations and data analysis}\label{sec:obs}
\subsection{XMM-Newton analysis}
\subsubsection{Data reduction}

\RXJ is the most luminous cluster in the sample of \citet{2015MNRAS.450.4248B}. From the ROSAT data, this cluster already appears to be one of the most luminous known at high redshifts with { an observer-frame luminosity of $L_{{\rm obs,}0.1-2.4\,{\rm keV}} = 20.3 \pm 6.2\times 10^{44}$\,erg/s (\citealp{2015MNRAS.450.4248B})}. They measure the spectroscopic redshift to be $z=0.902$ and their SZ data yields a mass of $M_{500}=(5.3\pm1.5)\times10^{14}{\, h_{70}^{-1}\rm M_\odot}$.

We analyze XMM-Newton observations of the cluster with ${\sim}15$\,ks cleaned exposure time (XMM-Newton observation IDs 0722530101 and 0722530201, PI of the joint XMM-Newton and HST program: T. Schrabback). The observations were performed in October and November 2013 (see Tab. \ref{tab:observation_xmm}) and were executed over the course of two revolutions, which we analyze simultaneously. 

Following the standard data reduction procedure\footnote{see \texttt{heasarc.gsfc.nasa.gov/docs/xmm/abc/}} using SAS version 14.0.0, we use the ODF data and apply {\it cifbuild} to catch up with the latest calibration and {\it odfingest} to update the ODF summary file with the necessary instrumental housekeeping information. Then we proceed by applying {\it emchain} and {\it epchain} (for MOS and PN detector, respectively) to create calibrated event files. 

On these calibrated files we apply the following filters for the event pattern of the triggered CCD pixels (the numbering is based to the ASCA GRADE selection) and the quality flag of the pixels: PATTERN $\leq 12$ for the MOS detectors, for PN PATTERN $= 0$; FLAG $= 0$ for both detectors. Because of anomalous features on CCD4 of MOS1, we additionally filter out events falling onto this chip. The CCD3 and CCD6 of MOS1 were damaged by micro meteorite events and the data of these detectors cannot be used.

In a next step we create light curves for both revolutions and all detectors in the energy range $0.3 - 10$\,keV. 
The observation in the second revolution shows strong flaring for a large fraction of the exposure time. We apply a $3\sigma$ clipping to all the light curves to filter the flared time intervals and inspected the light curves afterwards which then show no further hint of flaring.
This removes approximately half of the exposure time for the second observation (revolution 2546). 

To detect point sources in the field of view (FOV), we create images from the event files for all detectors in five energy bands between $0.2-12$\,keV. These images are provided in the task {\it edetect\_chain}. 

\renewcommand{\arraystretch}{1.2}
\begin{table}
\caption{Details of the XMM-Newton observation of \RXJdot.}           
\centering                                      
\begin{tabular}{c c c c p{1.3cm} c} 
Rev. & date & R.A. & Dec. & Cleaned exp. time & Filter\\ \hline \hline                      
2545 & Oct. 2013 & 182.512 & 49.926& 9.6\,ks & thick \\
2546 & Nov. 2013& 182.510& 49.924& 5.1\,ks & thick \\
\hline \hline                                             
\end{tabular}
\label{tab:observation_xmm}
\end{table}
\renewcommand{\arraystretch}{1}

\subsubsection{Spectral fitting}\label{sec:fitting}

An X-ray image of the cluster is shown in Fig. \ref{fig:RXJ}. We select three annular regions around the center and choose the region sizes such that we can achieve a S/Bkg ratio (i.e., counts$_{source}$/counts$_{bkg}$) of ${\sim}1$ in the outermost annulus and higher for the inner regions to avoid systematic biases. The final regions are $0'-0\farcm3, 0\farcm3-0\farcm8, $ and $0\farcm8-1\farcm3$. 
We fit the spectra of all annuli and for all detectors and the two observations simultaneously using the Cash-Statistic (cstat option) in XSPEC. For the cluster emission we use an absorbed APEC model with a column density from \citet{2013MNRAS.431..394W}, which also includes molecular hydrogen and the solar metal abundance table from \cite{2009ARA&A..47..481A}. We assume the same abundance in all annuli and thus link the corresponding model parameters.
The XMM-Newton point spread function (PSF) is ${\sim}17''$ HEW. We correct for the effect of photon mixing between different annuli due of the PSF as described in Sec. \ref{sec:PSF}.

From our HST data (Sec. \ref{sec:hst_results}) we estimate $R_{500} = 1\farcm8$  and therefore, for the estimation of the global cluster properties, we extract spectra in this region. 
For the analysis of such a high redshift cluster, the background treatment is crucial. The different background components are described in Sec. \ref{sec:QPB} and \ref{sec:XRBG} and we follow two approaches for the treatment of the background:

\begin{enumerate}

\item {\bf Background modeling}
One approach is to model all the different background components individually in the fitting procedure. These components are described in the following sections. We determine models for the quiescent particle background and the X-ray background and use them in the fitting of the cluster emission. We additionally introduce a power-law model to account for the residual soft proton emission, which is left over emission after the flare filtering. The index is linked for the two MOS detectors while the normalizations for each detector are independent. We use an energy range between $0.7-10$\,keV. The results of this approach can be found in Sec. \ref{sec:xmm_results}.

\item  {\bf Background subtraction}
The cluster has a small extent on the sky, thus we do not expect significant cluster emission beyond $R_{200} = 2.7'$ estimated from our HST data. For this reason we are able to subtract the full background from the spectra. To do so, we extract background spectra in an annulus between $3' - 5'$. This region lies completely on the MOS CCD1 chips which is important because the particle background shows strong variations between the different chips. Also for PN this region is close enough to the source extraction region to properly model the Ni and Cu lines. As for the first method, the energy range is $0.7-10$\,keV and the results of this procedure are described in Sec. \ref{sec:xmm_results}.

\end{enumerate}

\subsubsection{Quiescent particle background}\label{sec:QPB}
\begin{figure}[b]
\resizebox{\hsize}{!}{\includegraphics{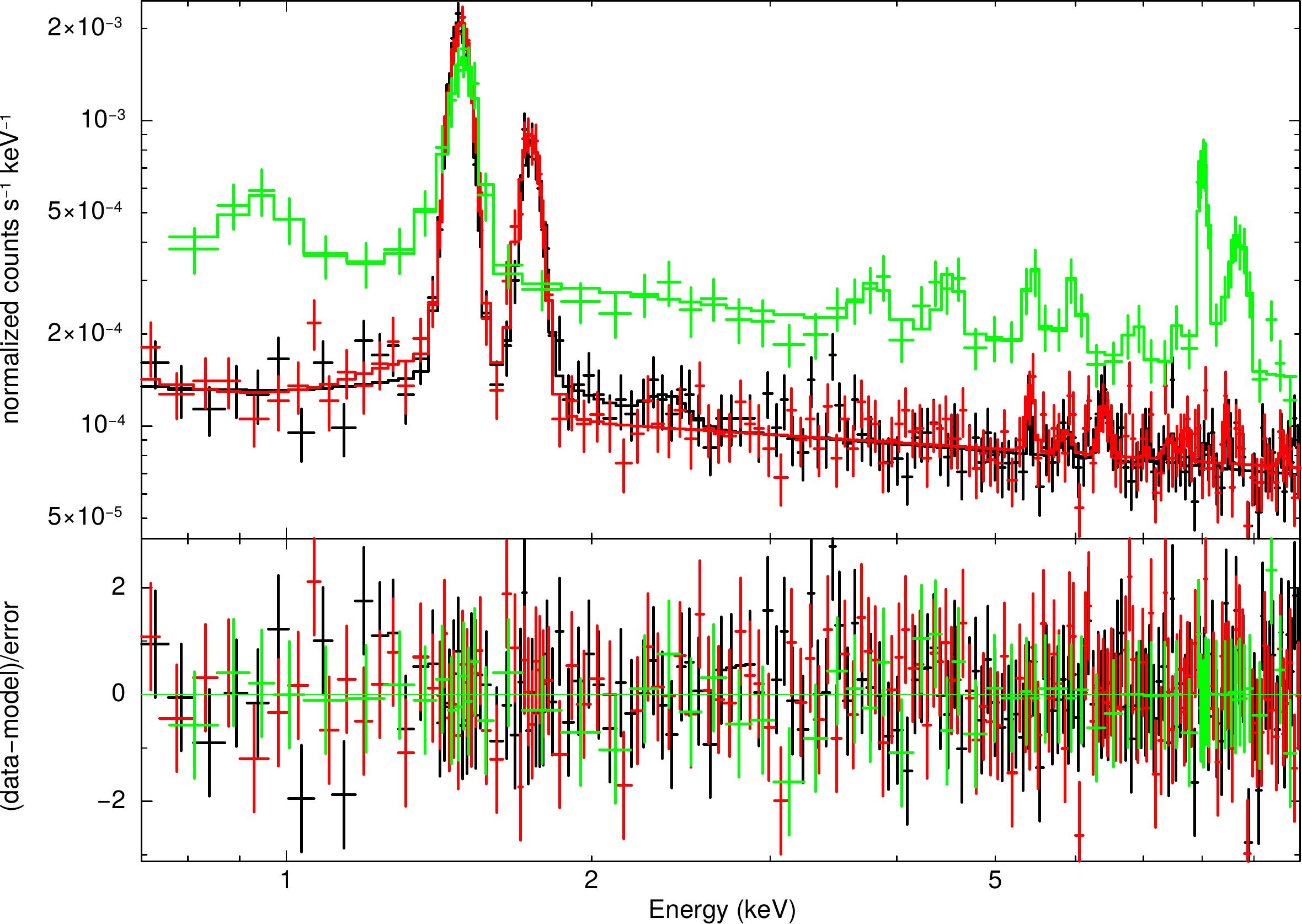}}
\caption{Spectra and best fit models of the QPB obtained from the filter-wheel-closed observations and extracted on the central chip in the region $0'-5'$ for MOS1 (black), MOS2 (red) and PN (green) and normalized to the extraction area.}
\label{fig:QPB_spectra}
\end{figure}

The quiescent particle background (QPB) is caused by highly energetic particles interacting with the detector and the surrounding material. It is composed of a continuum emission and fluorescent lines from various elements contained in the assembly of the satellite. XMM-Newton is equipped with a filter wheel system which can be used to measure the level of the QPB. When the filter is closed, only the high energy particles can penetrate the filter and a spectrum of the QPB can be obtained. We use merged event files of the filter-wheel-closed observations which are close to the time of the observation (revolution $2514-2597$ for the MOS detectors and $2467-2597$ for PN). 
The continuum part of the spectrum can be described by two power laws, while the fluorescent lines are modeled by Gaussians. The QPB varies for all detectors and with the position on the detector. Therefore, we fit the model in two regions -- from $0' - 5'$ (the source region, which lies completely on CCD1 for the MOS destectors) and from $7' - 12'$ (the region where we determine the X-ray background, see Sec. \ref{sec:XRBG}) -- for all detectors independently. For the QPB, diagonal responses are used in the fit and no ancillary response file (ARF) is applied as these particles do not suffer from instrumental effects such as vignetting. The spectra with the best fit models are shown in Fig. \ref{fig:QPB_spectra}. When fitting the cluster emission, the QPB normalizations of the power-law components and the Gaussian lines are allowed to vary separately by $\pm20\,\%$, due to possible spatial and temporal variations of the QPB.

\subsubsection{X-ray background}\label{sec:XRBG}
The X-ray background (XRBG) emission is caused by different sources: 1.) a local component and solar wind charge exchange,  2.) a component from the Milky Way halo plasma, and 3.) the superposition of the X-ray emission from distant AGNs causing a diffuse background (CXB). To model these background components we extract a spectrum in a region $7' - 12'$, where no cluster emission is expected. Additionally, ROSAT All-Sky-Survey data\footnote{obtained with the HEASARC X-ray background tool \texttt{heasarc.gsfc.nasa.gov/cgi-bin/Tools/xraybg/xraybg.pl}} are used to support the estimation of the background parameters at energies between $0.1-2.0$\,keV. 
The first XRBG component can be modeled using an APEC model with temperature and normalization as free fitting parameters. The redshift and the abundance are set to 0 and 1, respectively. The second component can be described by an absorbed APEC model. The superposition of AGN emission was analyzed by \citet{2004A&A...419..837D} and can be modeled by an absorbed power law with a photon index of 1.41. We accounted for the particle background in this annulus by using the previously determined model in Sec. \ref{sec:QPB} in the region $7'-12'$ with two floating multiplicative constants ($\pm 20\%$) for the continuum part and the fluorescent lines.
We additionally introduce a power-law model to account for the residual soft proton emission. Also for this model we use diagonal response matrices.

\begin{figure}
\resizebox{\hsize}{!}{\includegraphics{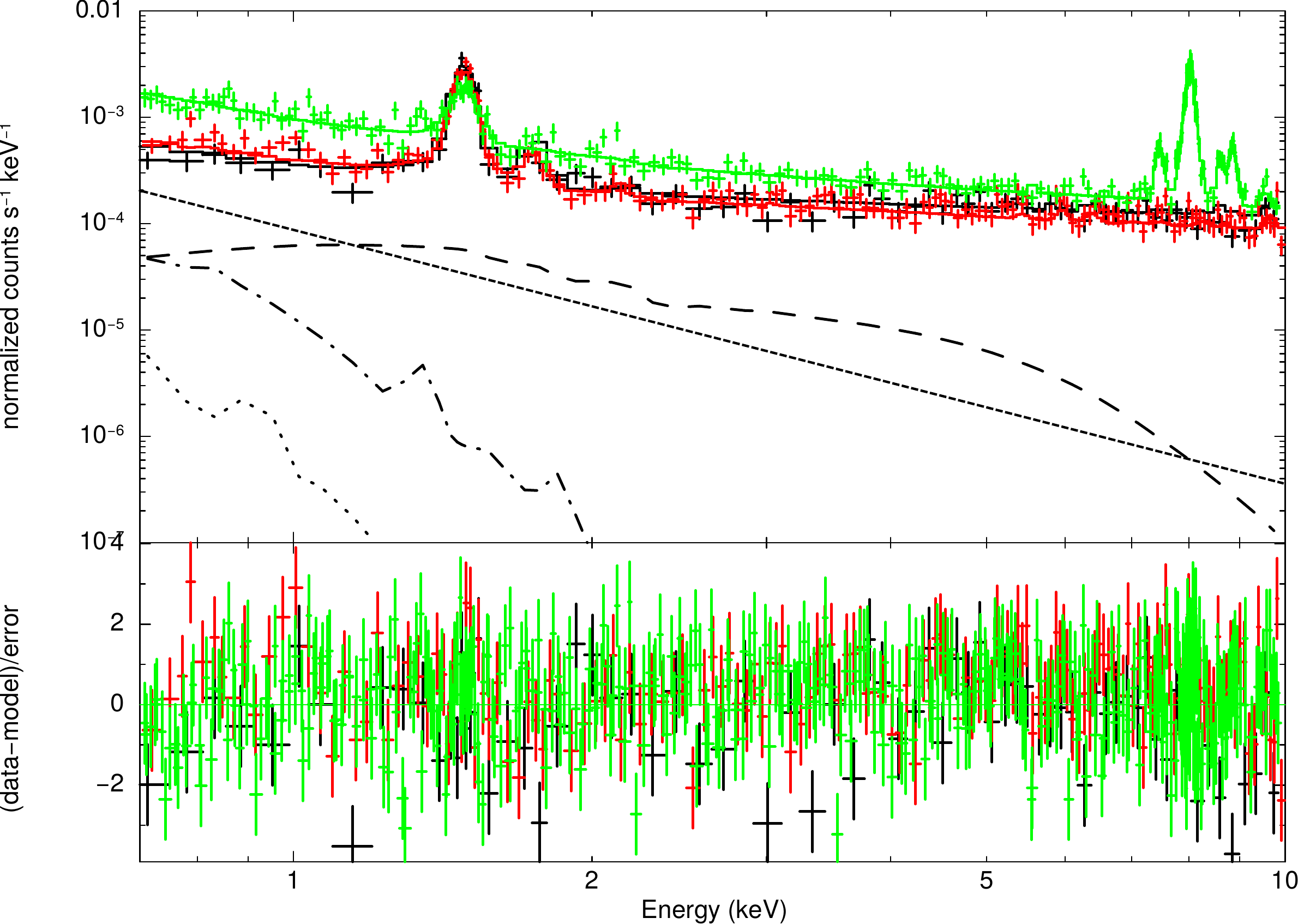}}
\caption{Spectra and best fit models for the XRBG + QPB for MOS1 (black), MOS2 (red), and PN (green) in the region $7'-12'$. The different components of the XRBG are shown as dotted, dash-dotted and dashed lines for the local, halo and CXB component, respectively. The power law component for the residual soft proton emission is shown as short-dashed line. For the spectra and models of the QPB see Fig. \ref{fig:QPB_spectra}.}
\label{fig:xrbg_spectra}
\end{figure}

The XRBG spectra and the best fit models for the different components are shown in Fig. \ref{fig:xrbg_spectra} for the off-axis region between $7'$ and $12'$.

\subsubsection{PSF correction}\label{sec:PSF}
The extent of the cluster on the sky is small; therefore, we have to choose annular region sizes which suffer from the PSF size of XMM-Newton. This causes ``mixing" of photons, i.e., photons originating from a certain region on the sky are detected in another region on the detector. This has an impact on the spectra and influences the measurements, especially the determination of the temperature profile. To avoid this we introduce a PSF correction. The XMM-Newton task {\it arfgen} allows us to calculate cross-region ARFs. Via these cross-region ARFs the effective area for the emission coming from one particular region, but detected in another, is estimated. These ARFs can then be used in the fitting process to account for the PSF effects. Therefore, we introduced additional absorbed APEC models for each combination of photon mixing (e.g., photons from region 1 on the sky but detected in region 2 on the detector, etc.). These models use the cross-region ARFs and the model parameters are linked to the parameters of the annulus the emission truly originates from, as described in the corresponding SAS-thread\footnote{\texttt{cosmos.esa.int/web/xmm-newton/sas-thread-esasspec}}. We neglect the PSF effects for the emission coming from the outermost annulus to the two inner annuli as the effective area for this mixing is close to zero.

\subsection{HST analysis}\label{sec:hst_analysis}
Here we perform a weak gravitational lensing analysis based on new Hubble
Space Telescope observations of \clusteranospace, obtained within the joint
XMM-Newton+HST program (HST program ID 13493).
Weak lensing measurements require accurate measurements of the shapes of
background galaxies well behind the cluster. 
Given the high redshift of \clusteranospace, typical weak lensing  background galaxies 
are at redshifts \mbox{$z\gtrsim 1.4$}. As most of them are unresolved in ground-based
seeing-limited data, HST observations are key for this study.
Specifically, we analyze
observations obtained with the {\it Advanced Camera for Surveys} (ACS) in
the F606W filter in a {\mbox{$2\times 2$} mosaic covering a \mbox{${\sim} 6\farcm5 \times
6\farcm6 $} area (corresponding to ${\sim}3.0\times3.1$\,Mpc$^2$)}, with integration times of 1.9\,ks
per pointing, each split into four exposures.

The data reduction and analysis is conducted with the same pipeline that was used 
for the weak lensing analysis of high-redshift galaxy clusters from the
South Pole Telescope Sunyaev-Zel'dovich Survey \citep{bleem15} presented in
\citet[][hereafter \citetalias{schrabback16}]{schrabback16}.
Therefore, we only summarize the main analysis steps here
 and refer the
reader to \citetalias{schrabback16} for further details.

For the ACS data reduction we employ basic calibrations from 
\texttt{CALACS}, the
  correction for charge-transfer
inefficiency from \citet{massey2014}, 
 \texttt{MultiDrizzle} \citep{koekemoer2003} for the cosmic ray removal and
 stacking,
and scripts for the image registration and improvement of masks from
\citet{schrabback2010}.

We detect objects using \texttt{Source Extractor} \citep{bertin1996} and
measure shapes using the  KSB+ formalism
\citep{kaiser1995,luppino1997,hoekstra1998} as implemented by
\citet{erben2001} with adaptions for HST measurements described in
\citet{schrabback07,schrabback2010}.
In particular, we apply a model for the
temporally and spatially varying HST PSF constructed
from a principal component analysis of ACS stellar field observations.
In order to estimate cluster masses from weak lensing, accurate knowledge of
the source redshift distribution is required. Here we follow  the approach from
\citetalias{schrabback16}, who first apply a  color selection to remove
cluster galaxies from the source sample, and then estimate the redshift
distribution based on CANDELS photometric redshift catalogs \citep{skelton14}, 
to which they apply consistent selection criteria, as used in the cluster fields,
and statistical corrections for photometric redshift outliers.

For the color selection we make use of additional $i$-band observations of \clustera obtained
with the Prime Focus Camera PFIP (Prime Focus Imaging Platform)
on the 
4.2\,m William Herschel Telescope (ID: W14AN004, PI: Hoekstra) on March 26, 2014.
These observations were taken with the new red-optimized RED+4
detector, which has an imaging area of \mbox{$4096\times 4112$} pixels, with
a pixel scale of 0\farcs27 and an \mbox{$18^\prime\times 18^\prime$} field
of view.
We reduce these data using \texttt{theli} \citep{erben05,schirmer13}, 
 co-adding exposures of a total integration time of $13.5$\,ks and reaching
 a  $5\sigma$ limit of 
\mbox{$i_\mathrm{WHT,lim}\simeq 25.8$} in circular
 apertures of $2^{\prime\prime}$, {with an image quality of \mbox{$2 r_\mathrm{f}=1\farcs2$}, where
$r_\mathrm{f}$ corresponds to the FLUX\_RADIUS parameter from
\texttt{Source Extractor}.} 
We use SDSS \citep{sdssdr13} for the photometric calibration and convolve
the ACS F606W imaging to the ground-based resolution to measure
\mbox{$V_\mathrm{606,con}-i_\mathrm{WHT}$} colors.
For galaxies at the cluster redshift the 4000\AA/Balmer break is located
within this filter pair. Therefore, by selecting very blue galaxies in this color, we
can cleanly remove the cluster  galaxies, while selecting the majority of the
\mbox{$z\gtrsim 1.4$} background sources carrying the lensing signal
 (see
\citetalias{schrabback16}).
To account for the increased scatter at faint magnitudes we apply a magnitude-dependent selection 
\mbox{$V_\mathrm{606,con}-i_\mathrm{WHT}<0.16$} (\mbox{$V_\mathrm{606,con}-i_\mathrm{WHT}<-0.04$})
for galaxies with magnitudes \mbox{$24<V_\mathrm{606}<25.5$}
(\mbox{$25.5<V_\mathrm{606}<26$}) measured in  0\farcs7 diameter apertures from the non-convolved
ACS images.
These cuts correspond to a color selection in the CANDELS catalogs of
\mbox{$V_{606}-I_{814}<0.2$} (\mbox{$V_{606}-I_{814}<0.0$}).
In order to select consistent galaxy populations between
the cluster field and the CANDELS catalogs we additionally apply consistent
lensing shape cuts and add photometric
scatter to the deeper CANDELS  catalogs as empirically estimated in
\citetalias{schrabback16}.
The depth of our final weak lensing catalog for \clustera\,
is mostly limited by the mediocre seeing conditions during the WHT observations, which
require us to substantially degrade the F606W images in the PSF matching
for the color measurements. 
As a result, we have to apply a rather stringent selection
\mbox{$V_\mathrm{606,auto}<25.8$}
based on the \texttt{Source Extractor} auto magnitude, which results in a final
galaxy number density of 9.6/arcmin$^2$, while the shape
catalog extends to \mbox{$V_\mathrm{606,auto}\simeq
  26.5$}.
We therefore recommend that future programs following a similar observing
strategy should ensure that
complementary
ground-based observations are conducted under good seeing conditions in order to fully exploit the statistical power
of the HST weak lensing shape catalogs.

{Taking the magnitude distribution and shape weights of our color-selected source catalog into account, we estimate an effective mean geometric lensing efficiency of \mbox{$\langle\beta\rangle=0.357\pm 0.009 (\mathrm{sys.}) \pm 0.025  (\mathrm{stat.})$} based on the CANDELS analysis (see \citetalias{schrabback16} for details).}

\section{Results}\label{sec:results}

\subsection{HST results} \label{sec:hst_results}
\begin{figure}
 \includegraphics[width=0.9\columnwidth]{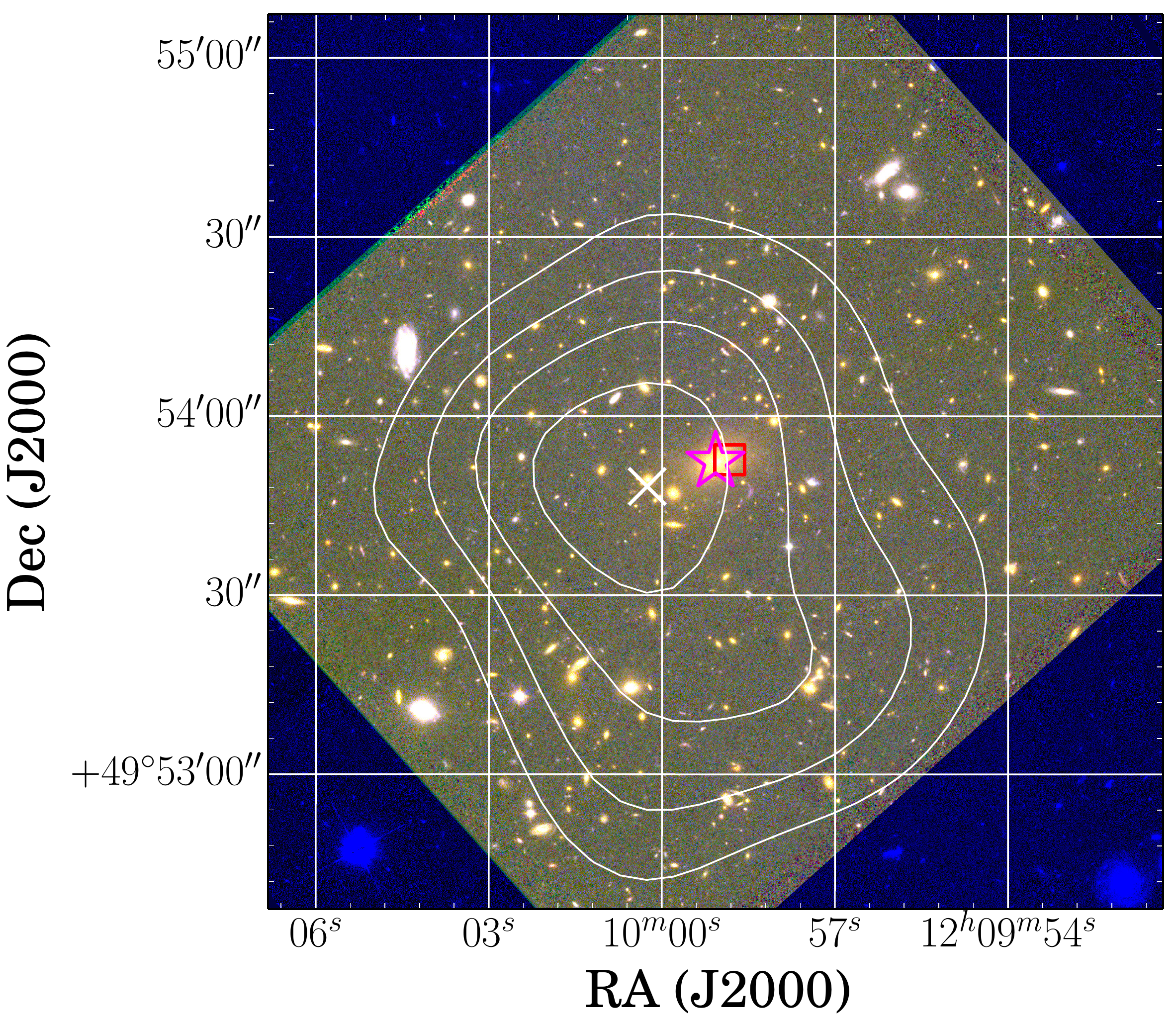}
 \caption{HST $2\farcm5 \times 2\farcm5$ color image of \clustera based on the
   ACS/WFC F606W (blue) and WFC3/IR F105W (green), and F140W (red) imaging.
The white contours indicate the signal-to-noise ratio of the weak lensing mass reconstruction,
starting at $2\sigma$ in steps of  $0.5\sigma$, with the cross marking the
peak position,
which is consistent with the X-ray peak (red square) and BCG position
(magenta star) within the uncertainty of 23$^{\prime\prime}$ in each direction.
\label{fig:massrecon}}
\end{figure}

In Fig.\thinspace\ref{fig:massrecon} we show contours of the weak
lensing mass
reconstruction of \clusteranospace, overlaid onto a color image from the
ACS/WFC F606W imaging and WFC3/IR imaging obtained in F105W (1.2\,ks)
and F140W (0.8\,ks). 
The reconstruction employs a Wiener filter \citep{mcinnes09,simon09}, as
further detailed in \citetalias{schrabback16}. Divided by
the r.m.s. image of the reconstructions of 500 noise fields, the contours
indicate the signal-to-noise ratio of the weak lensing mass reconstruction,
starting at $2\sigma$ in steps of  $0.5\sigma$.
The reconstruction peaks at \mbox{$\mathrm{R.A.}=$12:10:00.26},
\mbox{$\delta=$+49:53:48.2},
with a positional uncertainty of 23$^{\prime\prime}$ in each direction
(estimated by bootstrapping the source catalog), which makes it consistent
with the locations of the X-ray peak and the BCG at the  $1\sigma$ level.

\begin{figure}
 \includegraphics[width=0.9\columnwidth]{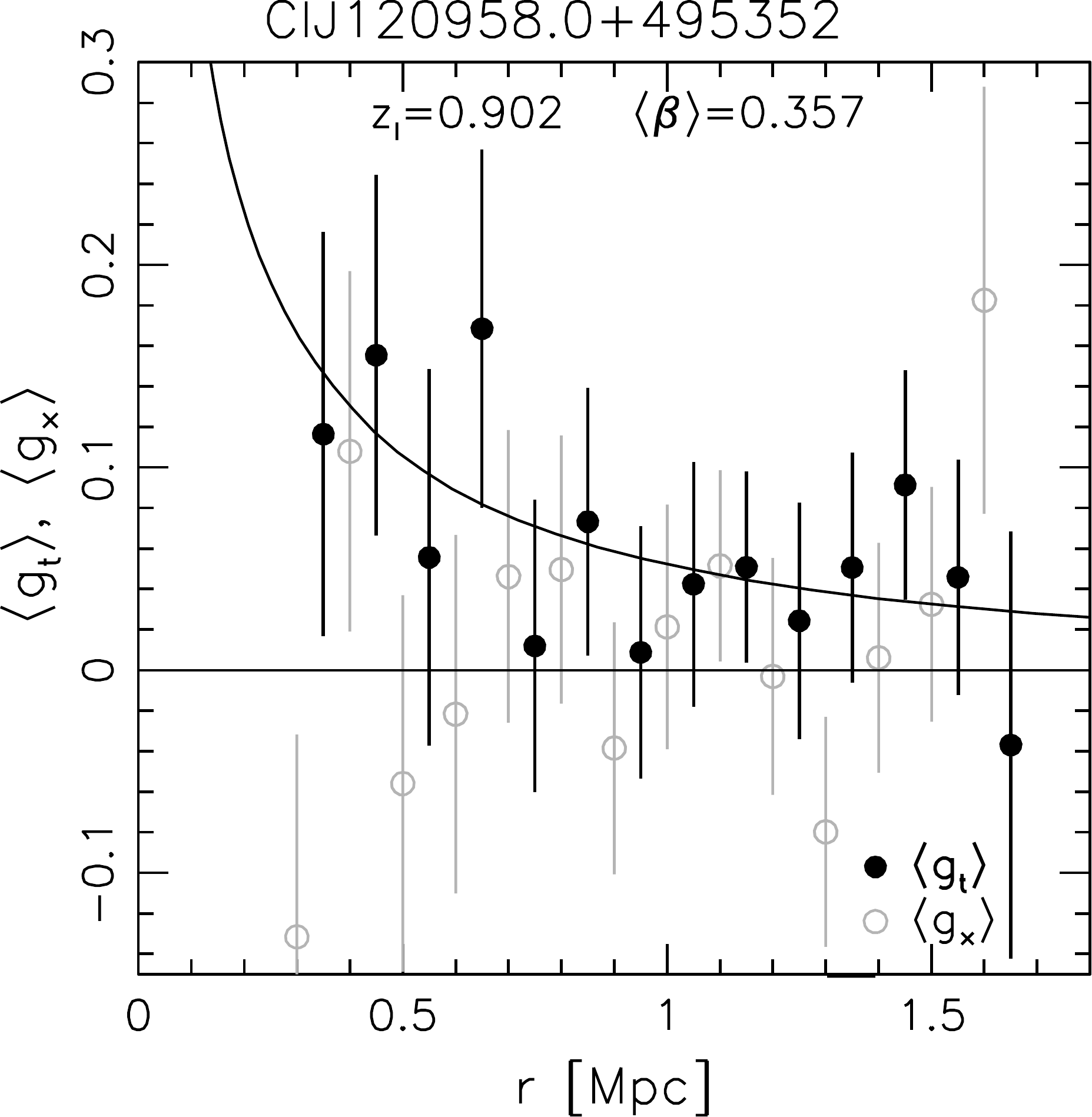}
 \caption{Tangential reduced shear profile (black solid circles) of
   \clusteranospace, measured around the 
X-ray peak. Here we  combine the profiles of four
 magnitude bins between \mbox{$24<V_\mathrm{606,aper}<26$}, as done
 in \citetalias{schrabback16}. The curve shows the
 corresponding best fitting NFW model prediction constrained by
 fitting the data within the range \mbox{$300\thinspace \mathrm{kpc}\le r \le 1.5
   \thinspace\mathrm{Mpc}$}, assuming 
the mass-concentration relation from \citet{diemer15}.
The gray open circles indicate the reduced cross-shear
component, which has been rotated by 45 degrees and constitutes a  test for systematics.
These points have been shifted by \mbox{$dr=-0.05$\thinspace Mpc}
 for clarity.
\label{fig:massplotdemo_profile}}
\end{figure}

Fig.\thinspace\ref{fig:massplotdemo_profile}
displays the measured tangential reduced shear profile of \clustera\, 
as a function of the projected separation from the X-ray peak,
combining measurements 
from all selected galaxies with  \mbox{$24<V_\mathrm{606,aper}<26$}, as done in
\citetalias{schrabback16}.
Fitting these measurements within the 
 radial range  \mbox{$300\thinspace \mathrm{kpc}\le r \le 1.5 \thinspace\mathrm{Mpc}$} 
assuming a model for a spherical NFW density profile 
according to
\citet{wright00}
and the mass-concentration relation from \citet{diemer15},
we constrain the cluster mass to 
\mbox{$M_\mathrm{500}/10^{14}{\rm M}_\odot=4.4^{+2.2}_{-2.0}(\mathrm{stat.})\pm
0.6  (\mathrm{sys.})$}
and
\mbox{$M_\mathrm{200}/10^{14}{\rm M}_\odot=6.5^{+3.0}_{-2.9}(\mathrm{stat.})\pm
 0.8 (\mathrm{sys.})$}.

Here we have corrected for a small expected bias  of 
$-7$\% ($-8$\%) for $M_\mathrm{500}$ ($M_\mathrm{200}$) caused by the simplistic
mass model, as estimated by
\citetalias{schrabback16} and further detailed in Applegate et al.~(in
prep.) using the analysis of simulated cluster weak lensing data.
Differing from  \citetalias{schrabback16} we assume negligible miscentering
for the bias correction, justified by the regular morphology of the cluster
and precise estimate of the X-ray cluster center.
The quoted statistical uncertainty includes shape noise, uncorrelated large-scale structure projections, and line-of-sight variations
in the source redshift distribution, while the systematic error estimate
takes shear calibration, redshift errors, and mass modeling uncertainties
into account  (see \citetalias{schrabback16} for details).
Here we have doubled the systematic mass modeling uncertainties used in
\citetalias{schrabback16} as we include somewhat smaller scales
in the fit{\footnote{In the analysis of simulated data we find that the mass biases increase by factors of \mbox{${\sim} 1.6-2.3$} when changing from the default lower limit  \mbox{$>500$ kpc} from \citetalias{schrabback16} to \mbox{$>300$ kpc} as employed here.
Following  \citetalias{schrabback16}, we estimate the residual uncertainty of the bias correction as a relative factor of the bias value. Accordingly, the uncertainty increases by approximately a factor of two.}}. 
When restricting the radial range in the fit to the more conservative range
\mbox{$500\thinspace \mathrm{kpc}\le r \le 1.5 \thinspace\mathrm{Mpc}$} from 
\citetalias{schrabback16}, the  
resulting constraints are
\mbox{$M_\mathrm{500}/10^{14}M_\odot=4.2^{+2.6}_{-2.3}(\mathrm{stat.})\pm
0.4  (\mathrm{sys.})$}
and
\mbox{$M_\mathrm{200}/10^{14}M_\odot=6.3^{+3.6}_{-3.4}(\mathrm{stat.})\pm
 0.6 (\mathrm{sys.})$}
with smaller expected and corrected biases of 3\% (5\%) for
$M_\mathrm{500}$ ($M_\mathrm{200}$) and smaller systematic uncertainties,
but increased statistical errors.

For the comparison to the X-ray measurements we additionally require weak
lensing mass estimates
for an overdensity \mbox{$\Delta=2500$}. 
When assuming the \citet{diemer15} mass-concentration relation 
and extrapolating the  bias corrections\footnote{This is necessary given that 
the analysis from \citetalias{schrabback16} 
as a function of $\log{\Delta}$ provides bias estimates
for \mbox{$\Delta=200$}  and \mbox{$\Delta=500$} only, as masses
\mbox{$M_\mathrm{2500}$} are not available for the simulations used to
derive the bias values. We do propagate the statistical uncertainty of this
extrapolation, but note that it is negligible compared to the statistical
uncertainty of the mass constraints for \clusteranospace.},
the weak lensing mass constraints correspond to
\mbox{$M_\mathrm{2500}/10^{14}M_\odot=1.7^{+0.9}_{-0.8}(\mathrm{stat.})\pm 0.2 (\mathrm{sys.})$}
when including measurements from scales \mbox{$300\thinspace \mathrm{kpc}\le
  r \le 1.5 \thinspace\mathrm{Mpc}$},
and 
\mbox{$M_\mathrm{2500}/10^{14}M_\odot=1.6^{+1.0}_{-0.9}(\mathrm{stat.})\pm 0.2 (\mathrm{sys.})$}
when restricting the analysis to scales \mbox{$500\thinspace \mathrm{kpc}\le
  r \le 1.5 \thinspace\mathrm{Mpc}$}.

We expect that our mass estimation procedure is unbiased within the quoted
systematic uncertainties for a random population of massive clusters. For an
individual cluster like the one studied here, deviations in the density profile from
the assumed NFW profile with a concentration from the \citet{diemer15} 
mass-concentration relation lead to additional scatter in the mass
estimates. 
To estimate the order of magnitude of this effect we repeat the mass fits
for scales \mbox{$300\thinspace \mathrm{kpc}\le
  r \le 1.5 \thinspace\mathrm{Mpc}$}
using different concentrations. {
Based on simulations, \citet{2008MNRAS.390L..64D} find that the scatter around the median concentration is approximately lognormal with $\sigma(\log_{10}c_{200})=0.11$ for relaxed clusters.
Approximately matching the expected $1\sigma$ limits, fixed concentrations \mbox{$c_\mathrm{200}=3.0$} (\mbox{$c_\mathrm{200}=5.0$}) change the best fit mass constraints for \mbox{$M_\mathrm{200}, M_\mathrm{500}, M_\mathrm{2500}$} by
 \mbox{$+11\%, +6\%, -9\%$} (\mbox{$-11\%, -5\%, +11\%$}) compared to the default analysis using the \citet{diemer15}
 mass-concentration relation which yields a concentration of
 \mbox{$c_\mathrm{200}=3.7$} at the best fitting mass.}
These variations are negligible compared to the statistical uncertainties 
of the study presented here. It should be noted that this analysis assumes spherical cluster models, which can lead to extra scatter due to triaxiality when comparing to X-ray results.

{In Fig.\thinspace\ref{fig:massrecon}, the signal-to-noise ratio contours of the mass reconstruction appear to be slightly elliptical, extending towards the south-southwest, {which is tentatively in agreement with the location of some apparent early-type cluster galaxies}.
To investigate whether this elliptical shape is actually significant, we estimate the shape of the mass peak using \texttt{Source Extractor} both for the actual mass reconstruction and the reconstructions originating from the bootstrap-resampled catalogs.
Using the \texttt{Source Extractor} estimates of the semi-major and semi-minor axes $a$ and $b$, as well as the position angle $\phi$ measured towards the north from west, we compute complex ellipticities \mbox{$e=e_1+\mathrm{i}\, e_2=|e| \, \mathrm{e}^{2\mathrm{i}\phi}$} with \mbox{$|e|=(a-b)/(a+b)$}, as employed in weak lensing notation \citep[e.g.,][]{bartelmann01}.
Using the dispersion of the estimates from the boostrapped samples as errors, our resulting estimate  \mbox{$e=(-0.05\pm 0.18)+\mathrm{i}\, (-0.06\pm 0.16)$} is consistent with a round mass distribution (\mbox{$e=0$}). Hence, the apparent elliptical shape is not significant.}

\subsection{XMM-Newton results}\label{sec:xmm_results}
\subsubsection{Global cluster properties}
The global properties for both methods of the treatment of the background are summarized in Tab. \ref{tab:global_props}. 
The overall properties agree well between the two methods. 

The rest-frame luminosity of the cluster in the $0.1-2.4$\,keV band is $L_X = (13.4_{-1.0}^{+1.2})\times10^{44}$\,erg/s and $L_X =(13.7_{-0.5}^{+0.5})\times10^{44}$\,erg/s, for background-modeling and background-subtraction method, respectively, estimated from the spectral fit. It is thus comparable to the most X-ray luminous MACS clusters, but at even higher redshift. These values are also in very good agreement with the findings by \citet{2015MNRAS.450.4248B} { after applying a K-correction}.

\renewcommand{\arraystretch}{1.2}
\begin{table}
\caption{Global cluster properties between $0'< R <1\farcm8$}           
\centering                                      
\begin{tabular}{c p{1.5cm} | p{1.5cm}} 
& background-modeling & background-subtraction \\
\hline \hline                      
$T$ [keV] &$ 9.04 _{ -1.88 }^{+ 1.38 }$&$ 8.84 _{ -0.71 }^{+ 0.97 }$\\
$Z$ [$Z_\odot$] &$ 0.35 _{ -0.18 }^{+ 0.20 }$&$ 0.46 _{ -0.17 }^{+ 0.19 }$\\
norm$^1$ &$ 18.95 _{ -1.28 }^{+ 1.32 }$&$ 19.09 _{ -0.73 }^{+ 0.72 }$\\

\hline \hline                                             
\end{tabular}
\label{tab:global_props}

\begin{minipage}{\columnwidth}
\vspace{0.2cm}
{\tiny $^1 {\rm norm}=\frac{10^{-18}}{4\pi[D_A(1+z)]^2}\int n_{\rm e}n_{\rm H}{\rm d}V\,{\rm cm}^{-5}$ with $D_A$ being the angular diameter distance to the source.}
\end{minipage}

\end{table}
\renewcommand{\arraystretch}{1}

\subsubsection{Temperature and density}\label{sec:tempdens}
\begin{figure}
\resizebox{\hsize}{!}{\includegraphics{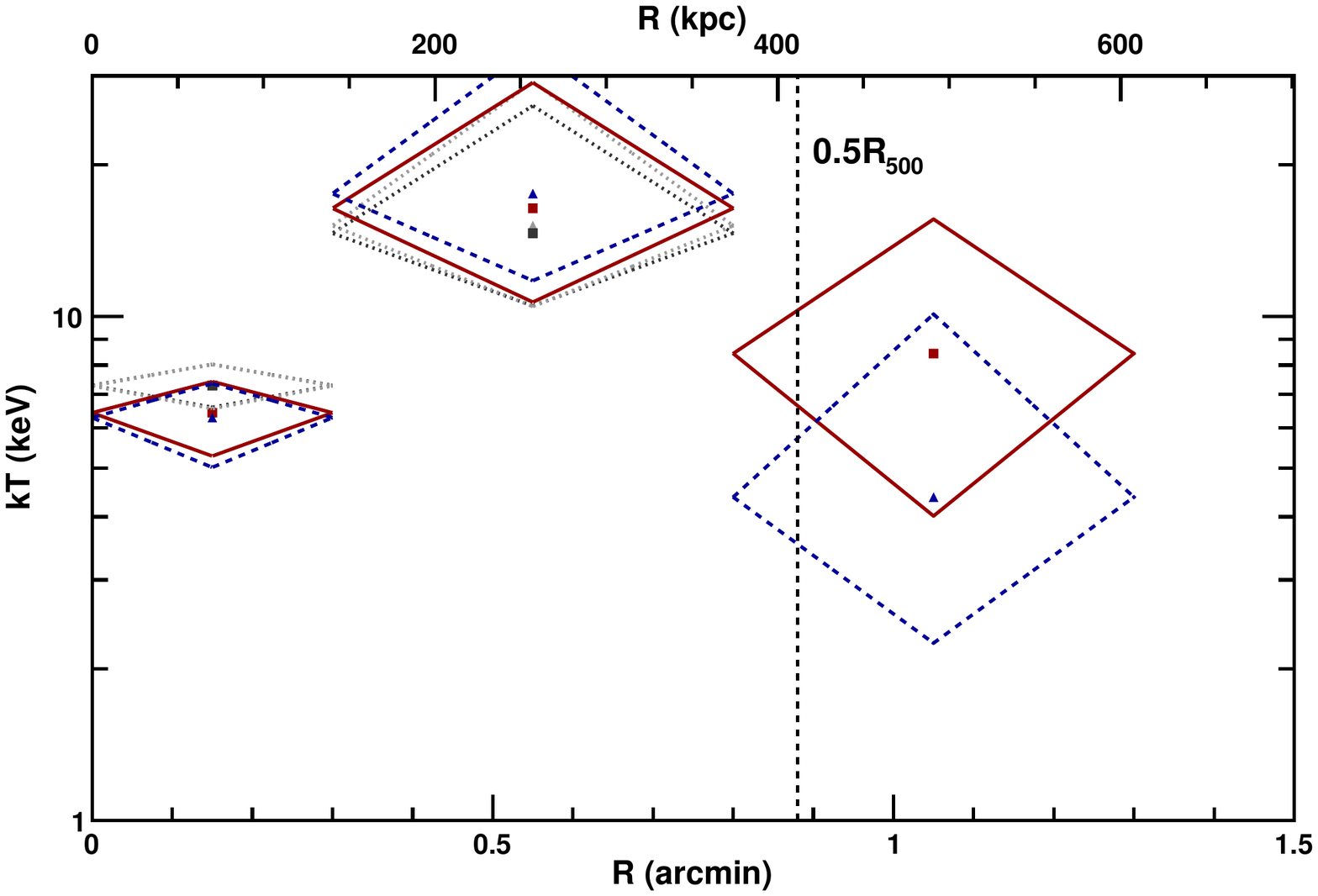}}
\caption{Deprojected and PSF-corrected temperature profile of \RXJdot. Red (dark gray) solid diamonds show the deprojected (projected) result using the background-subtraction method. Blue (light gray) dashed diamonds corresponds to the background-modeling method.}
\label{fig:temp_profile}
\end{figure}

We compare the results for the two approaches of the background treatment for temperature and density profile. Fig. \ref{fig:temp_profile} shows the temperature profile of \RXJ for both approaches, and Tab. \ref{tab:fitresults} gives the results.

\renewcommand{\arraystretch}{1.2}
\begin{table}
\caption{Fit results for the three radial bins for both methods of background treatment. The abundance is linked between all annuli.}           
\centering                                      
\begin{tabular}{c c c c} 
& $0'-0\farcm3$ & $0\farcm3-0\farcm8$ & $0\farcm8-1\farcm3$ \\
\hline \hline   
\multicolumn{4}{c}{background-modeling}\\\hline
$T$ [keV] &$ 7.28 _{ -0.72 }^{+ 0.75 }$&$ 15.13 _{ -4.67 }^{+ 14.04 }$&$ 4.38 _{ -2.13 }^{+ 5.72 }$\\
$Z$ [$Z_\odot$]&\multicolumn{3}{c}{\Vhrulefill ~$0.25 _{ -0.14 }^{+ 0.16 }$ ~\Vhrulefill}\\
norm$^1$ &$ 11.38 _{ -0.49 }^{+ 0.58 }$&$ 5.40 _{ -0.46 }^{+ 0.44 }$&$ 2.08 _{ -0.46 }^{+ 0.86 }$\\\hline

\multicolumn{4}{c}{background-subtraction}\\\hline
$T$ [keV] &$ 7.29 _{ -0.69 }^{+ 0.74 }$&$ 14.61 _{ -4.13 }^{+ 11.55 }$&$ 8.43 _{ -4.42 }^{+ 7.15 }$\\
$Z$ [$Z_\odot$]&\multicolumn{3}{c}{\Vhrulefill ~$0.32_{-0.15}^{+0.17}$ ~\Vhrulefill}\\
norm$^1$ &$ 11.24 _{ -0.51 }^{+ 0.53 }$&$ 5.34 _{ -0.43 }^{+ 0.44 }$&$ 1.82 _{ -0.28 }^{+ 0.47 }$\\

\hline \hline                                             
\end{tabular}
\label{tab:fitresults}
\begin{minipage}{\columnwidth}
\vspace{0.2cm}
{\tiny $^1 {\rm norm}=\frac{10^{-18}}{4\pi[D_A(1+z)]^2}\int n_{\rm e}n_{\rm H}{\rm d}V\,{\rm cm}^{-5}$ with $D_A$ being the angular diameter distance to the source.}
\end{minipage}

\end{table}
\renewcommand{\arraystretch}{1}

Overall we see a very good agreement between the two different background methods. 
The temperature of the central bin is well constrained in both cases and both profiles show a good indication of a cool core. This makes \RXJ one of only a few such objects known at high redshifts.
The upper uncertainties in the outer two bins are large which is mainly related to the correlation between the parameters due to the PSF correction and the limited statistics. Even if no PSF correction is applied, the cool core remains and the uncertainty of the second temperature decreases by a factor of ${\sim}5$ and of the outermost temperature by a factor of ${\sim}2$.

We determine the gas density profile using the PSF-corrected normalizations of the APEC model, which is defined as

\begin{equation}
\label{eq:norm}
N_i = \frac{10^{-14}}{4\pi D_{\rm A}^2(1+z)^2}\int_{V_i} n_{\rm e}(R)n_{\rm H}(R)\,{\rm d}V ,
\end{equation}
where $i$ corresponds to the $i$th annulus from the center and $D_{\rm A}$ is the angular diameter distance to the source. The volume along the line of sight $V_i$ is the corresponding cylindrical cut through a sphere with inner and outer radii of the $i$th annulus. We adopt $n_{\rm e} = 1.17 n_{\rm H}$. 
Due to the small extent of the cluster, there is only limited radial resolution. Therefore, we perform a simple deprojection method following \citet{2002MNRAS.331..635E}.

\begin{figure}
\resizebox{\hsize}{!}{\includegraphics{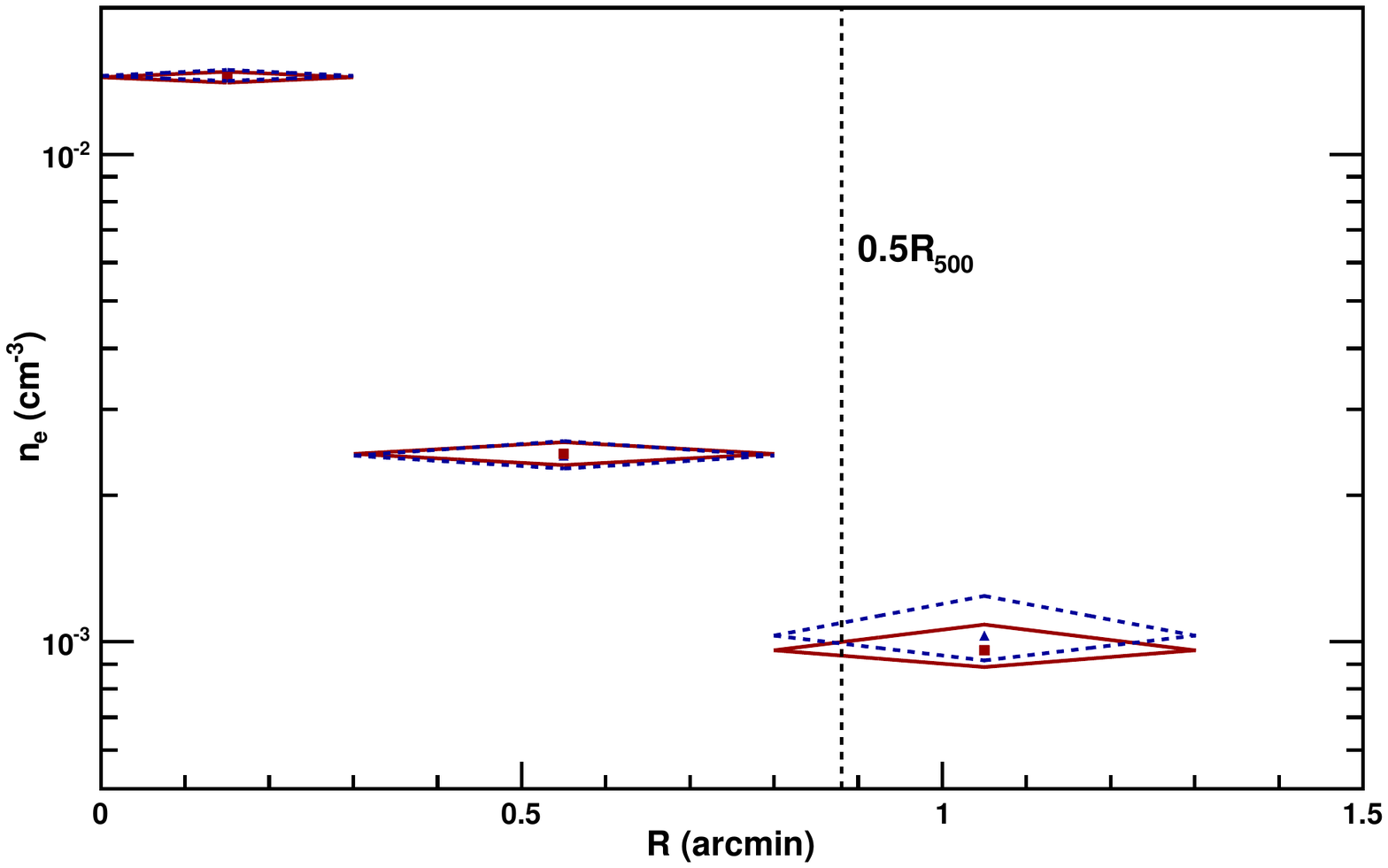}}
\caption{Deprojected and PSF-corrected electron density profile of \RXJdot. Red solid diamonds show the result using the background-subtraction method. Blue dashed diamonds correspond to the background-modeling method. The width of the diamonds corresponds to the radial bin size.}
\label{fig:dens_profile}
\end{figure}

The emission integral ($EI_i$) and temperature ($T_i$) in ring $i$ are given by
\begin{align}
 EI_i &= \sum_{j=i}^{N} n_{\rm e}n_{\rm H} V_{i,j} \\
 T_i &=  \frac{\sum_{j=i}^{N} \epsilon_j V_{i,j} T_{j}}{\sum_{j=i}^{N} \epsilon_j V_{i,j}},
\end{align}
with $V_{i,j}$ being the volume of the cylindrical cut corresponding to ring $i$ through spherical shell $j$, $n_{\rm e}$ and $n_{\rm H}$ the electron and proton density, and $\epsilon$ the emissivity. By subtracting the contribution of the overlying shells in each annulus, we determine the deprojected electron density profiles for both background treatment methods shown in Fig. \ref{fig:dens_profile}. As for the temperature, the two density profiles agree very well showing that our background treatment works well in both cases.

As an additional test for the background-subtraction method we choose an even larger inner radius of the background region ($4'-5'$) and repeat the analysis. We find only marginal differences and thus conclude that no significant cluster emission is present in the background-region.

As can be seen in Fig. \ref{fig:RXJ}, we detect a point source close to the center of the cluster. To investigate the impact of the point source, we increase the exclusion radius around this source by 50\% and repeat the fit. Due to the lowered statistics, the uncertainties clearly increase but we find no significant impact compared to the nominal values.

\subsubsection{Gas mass fraction}\label{sec:CIJ_results_fgas}

From the gas mass profile and the total mass $M_{\rm tot}(<R)$ inside a given radius $R$, the gas mass fraction can be obtained:

\begin{equation}
        f_{\rm gas}(<R) = \frac{M_{\rm gas}(<R)}{M_{\rm tot}(<R)} .                                                                                                                         
\end{equation}

We note that, given the limited XMM-Newton spatial resolution, a very robust determination of the total mass from the hydrostatic equation is difficult as this would require high spatial resolution measurements of the density and temperature profile. Therefore, we use the total mass based on our weak lensing HST estimates and the corresponding $R_{2500}$ (see Sec. \ref{sec:hst_results}). As a cross-check, we also determine the gas-mass fraction using the $L_X-M_{2500}$ relation obtained by \citet{2007MNRAS.379..317H} for the total mass.

The HST results yield $M_\mathrm{2500}/10^{14}M_\odot=1.7^{+0.9}_{-0.8}(\mathrm{stat.})\pm 0.2 (\mathrm{sys.})$. For the estimation of $f_{\rm gas}$, we include an additional 30\% triaxiality/projection uncertainty and a 10\% uncertainty from the mass-concentration relation on $M_\mathrm{2500}$. From 10000 Monte Carlo (MC) realizations of $M_\mathrm{2500}$, we estimate $R_{2500}=0\farcm75_{-0.20}^{+0.13}$ and for each realization the gas mass within the corresponding $R_{2500}$, assuming a constant density in each shell. This yields $M_{\rm gas,{2500}} = (1.64_{-0.67}^{+0.53})\times10^{13}$\,M$_\odot$ and $M_{\rm gas,{2500}} = (1.63_{-0.67}^{+0.53})\times10^{13}$\,M$_\odot$ for the background-subtraction and background-modeling method, respectively, which are in very good agreement. 
Combining these results, we estimate
$f_{{\rm gas},2500} = 0.10_{-0.02}^{+0.03}$ for both methods. We note that through this procedure the given uncertainties on $M_\mathrm{2500}$, $M_{\rm gas,{2500}}$, and $R_{2500}$ are, on the one hand, correlated and, on the other hand, the assumption of constant density in each shell is only a rough approximation, which is why the uncertainty on $f_{{\rm gas},2500}$ is lower than naively expected. A more general estimate is obtained by using a beta-model for the density profile and following the same procedure as described above. We fix the core radius to a typical value of $R_c = 0.15\times R_{500}$ and assume a slope of $\beta = 2/3$ (as also used in, e.g., \citealp{2016A&A...592A...2P}), but including 15\% scatter on the latter. $R_{500}$ is estimated from our HST results. This yields $f_{\rm gas} = 0.11_{-0.03}^{+0.06}$ for both background methods. Yet another approach is to estimate $f_{\rm gas}$ and its uncertainties at a {\it fixed} radius (i.e., assuming that the true $R_{2500}$ is known), in which case the uncertainties on $M_\mathrm{2500}$ and $M_\mathrm{gas}$ are uncorrelated and directly propagate onto $f_{\rm gas}$, which then yields $f_{\rm gas} = 0.11_{-0.05}^{+0.12}$. Here, we take the result using the beta-model as default.

\citet{2007MNRAS.379..317H} estimated the $L_X-M_{2500}$ relation for a galaxy cluster sample of 20 X-ray luminous objects at intermediate redshifts up to $z{\sim}0.6$. They find a slope consistent with the one from \citet{2009A&A...498..361P}, which is also used in the redshift evolution study of \citet{2011A&A...535A...4R} and also consistent with the (inverted) slope of \citet{2007ApJ...668..772M} who assumed self-similar evolution.
Using the relation from \citet{2007MNRAS.379..317H} and assuming 30\% intrinsic scatter, we find $M_{2500} =(1.31_{-0.29}^{+0.31})\times10^{14}$M$_{\odot}$ for the background-subtraction method and $M_{2500} =(1.30_{-0.30}^{+0.32})\times10^{14}$M$_{\odot}$ for the background-modeling method and (using the corresponding $R_{2500}$) $M_{\rm gas,{2500}} = (1.34_{-0.25}^{+0.27})\times10^{13}$\,M$_\odot$ and  $M_{\rm gas,{2500}} = (1.33_{-0.30}^{+0.32})\times10^{13}$\,M$_\odot$, respectively. This yields
$f_{{\rm gas},2500} = 0.10\pm0.02$ for both background methods, and is in very good agreement with our previous findings using the weak lensing mass.

\subsection{Cooling time}
To estimate the cooling time, we further reduced the size of the central region to $0\farcm2$ corresponding to ${\sim}100$\,kpc and performed the same PSF correction and deprojection method as described above. The cooling time is given by (\citealp{2010A&A...513A..37H})

\begin{equation}
 t_{\rm cool} = \frac{3(n_{\rm e} + n_{\rm i})k_{\rm B}T}{2n_{\rm e}n_{\rm H}\Lambda(T,Z)},
\end{equation}
where $n_{\rm i}$ is the ion density and $\Lambda(T,Z)$ the cooling function. Within $100$\,kpc we find $n_{\rm e} = (2.09_{-0.08}^{+0.10})\times 10^{-2}$\,cm$^{-3}$ and $T = 4.0_{-1.5}^{+1.3}$\,keV.
This yields a short cooling time for \RXJ within $100$\,kpc of $t_{\rm cool}~=~2.8\pm 0.5$\,Gyr for the background subtraction method and $t_{\rm cool} = 2.9\pm0.4$\,Gyr for the background modeling method. \citet{2010A&A...513A..37H} studied the cool cores for a local sample of 64 clusters within $0.4\%R_{500}$ with Chandra. According to their findings, \RXJ belongs to the weak cool core clusters; however, it should be taken into account that the radius, in which they determine the cooling time, is much smaller than what is possible for \RXJ and, presumably, within this radius the cooling time would be even lower, possibly resulting in a strong cool core classification.

\section{Discussion and conclusions}\label{sec:discussion}
Our results show that \RXJdot, according to \citet{2015A&A...581A..14P}, belongs to the most luminous galaxy clusters known at $z{\sim}0.9$. Compared to the total mass estimate from \citet{2015MNRAS.450.4248B} of $M_{500}=(5.3\pm1.5)\times10^{14}{\, h_{70}^{-1}\rm M_\odot}$, we find a slightly lower value from our weak lensing analysis of \mbox{$M_\mathrm{500}/10^{14}M_\odot=4.4^{+2.2}_{-2.0}(\mathrm{stat.})\pm0.6  (\mathrm{sys.})$}, which is, however, compatible within the uncertainties. 

As discussed in, e.g., \citet{2009MNRAS.398.1698S} and \citet{2012ApJ...761..183S} there is a tight correlation between the dynamical state of the cluster and the presence and strength of a cool core. 
We find strong indications for the presence of a cool core, and the two different approaches for the background handling yield similar results which gives us confidence in our treatment of the background. The temperature profile shows a clear drop towards the center and the cooling time within $100$\,kpc is short with $t_{\rm cool}~=~2.8\pm 0.5$\,Gyr and $t_{\rm cool} = 2.9\pm0.4$\,Gyr for the  background-subtraction and background-modeling method, respectively.
Another indicator for the morphological state is the offset between the BCG and the X-ray emission peak (see, e.g., \citealp{2016MNRAS.457.4515R}, \citealp{2013ApJ...767..116M}, \citealp{2010A&A...513A..37H}). \citet{2016MNRAS.457.4515R} define a relaxed cluster by an offset smaller than $0.02R_{500}$.
For \RXJ the offset is about $2''$ (${\sim}15$\,kpc) corresponding to $0.015R_{500}$ (using the BCG position given in \citealp{2015MNRAS.450.4248B}, see also Fig. \ref{fig:massrecon}), which is another indication for the relaxed nature of the system. Our HST weak lensing study also shows that the mass reconstruction peak is compatible with the BCG position and the X-ray peak within $1\sigma$. {As investigated in Sec.~\ref{sec:hst_results}, the apparent elliptical shape of the lensing mass reconstruction is not significant. Hence, the results are consistent with a round mass distribution.}

In a bottom-up scenario for structure formation, massive cool core systems should be extremely rare at high redshifts. Their gas mass fractions should not depend on the cosmological model. However, the apparent evolution varies for different assumed cosmologies. 
Previous measurements from \citet{2008MNRAS.383..879A} and \citet{2014MNRAS.440.2077M} show that their data are in good agreement with the standard cosmological model, showing a flat behavior of $f_{\rm gas}$ with redshift. However, these data only contain a few objects at very high redshifts. Therefore, clusters like \RXJ are valuable objects for cosmology.

 We obtain a gas mass fraction of $f_{\rm gas,2500} = 0.11_{-0.03}^{+0.06}$, which is consistent with the result from \citet{2008MNRAS.383..879A} for their full cluster sample and also consistent with the assumed $\Lambda$CDM cosmology ($\Omega_{\rm m}=0.3$, $h=0.7$). 
We performed several tests, i.e., we used an $L_X-M_{2500}$ scaling relation for the total mass and tested the assumption of constant density in each shell, to verify this result and find very good agreement.
 \citet{2014MNRAS.440.2077M} measured the gas mass fraction in an annulus from $0.8R_{2500}<R<1.2R_{2500}$ excluding the core of the clusters to minimize gas depletion uncertainties and intrinsic scatter in the inner part. They find typical $f_{\rm gas}$ values between $0.10-0.12$ and are thus consistent with our findings and \citet{2008MNRAS.383..879A}. 
 
\citet{2011A&A...535A...4R} studied the evolution of cluster scaling relations up to redshift $1.5$. They use the relations from \citet{2009A&A...498..361P} for the local clusters and obtain a bias-corrected evolution factor. Testing this $L_X- T$ scaling relation with our estimated global gas temperature yields a luminosity that is about 40\% smaller than our measured value. This result is, at least partially, expected due to the presence of a cool core. However, the uncertainties solely due to the uncertainties of the slope and normalization of the scaling relation (assuming they are uncorrelated) are already large ($\gtrsim40$\%).

The cluster \RXJ is interesting not only with respect to cosmology, but also to its astrophysics. At redshift $0.9$ the time span for this massive object to form a cool core is very short. As XMM-Newton is not able to fully resolve the core structure, we aim for higher spatial resolution data in a future project to robustly determine the X-ray hydrostatic mass and to perform a detailed study of the core properties.

\begin{acknowledgements}
 This work is based on joint observations made with the NASA/ESA {\it Hubble Space Telescope}, using imaging data from program 13493 (PI: Schrabback) and XMM-Newton data (IDs 0722530101 and 0722530201), as well as WHT data (ID W14AN004, PI: Hoekstra). ST and TS acknowledge support from the German Federal Ministry of Economics and Technology (BMWi) provided through DLR under projects 50 OR 1210,  50 OR 1308,  50 OR 1407, and 50 OR 1610. ST and THR acknowledge support by the German Research Association (DFG) through grant RE 1462/6 and the Transregio 33 ``The Dark Universe'' sub-project B18. ST also acknowledges support from the Bonn-Cologne Graduate School of Physics and Astronomy. LL acknowledges support from the Chandra X-ray Observatory grant GO3-14130B and from the Chandra X-ray Center through NASA contract NAS8-03060. Support for Program number GO-13493 was provided by NASA through a grant from the Space Telescope Science Institute, which is operated by the Association of Universities for Research in Astronomy, Incorporated, under NASA contract NAS5-26555.
\end{acknowledgements}

\bibliographystyle{aa} 
\bibliography{/users/thoelken/thoelken/UGC03957/Paper/accepted/bibtex,/users/thoelken/thoelken/ClGJ120959/Paper/oir}

\end{document}